%% file: kitzbichler.tex
\documentclass[useAMS,usenatbib,usegraphicx]{mn2e}

\usepackage{times}
\usepackage{amsfonts}
\usepackage{amssymb}
\usepackage{nicefrac}

\newcommand{\fig}{Fig.\,}

\newcommand{\eq}{Eqn.\,}
\newcommand{\tabl}{Table\,}
\newcommand{\secref}{Section\,\ref}

\newcommand{\kms}{\mbox{\,km\,s$^{-1}$}}

\newcommand{\kpchh}{\mbox{\,kpc$/h$}}
\newcommand{\kpch}{\mbox{\,$h^{-1}$kpc}}
\newcommand{\mpch}{\mbox{\,$h^{-1}$Mpc}}
\newcommand{\myrh}{\mbox{\,$h^{-1}$Myr}}
\newcommand{\msolh}{\mbox{\,$h^{-1}$M$_\odot$}}

\defcitealias{Croton2006}{Paper\,I}
\defcitealias{Delucia2006b}{Paper\,II}
\defcitealias{Kitzbichler2007}{Paper\,III}

\begin{document}

\title[Close galaxy pairs and merger rates]{A calibration of the relation
between the abundance of close galaxy pairs and the rate of galaxy mergers}
\author[Kitzbichler et al.]{M.~G.~Kitzbichler\thanks{E-mail:~mgk@mpa-garching.mpg.de} and S.~D.~M.~White\\ 
Max-Planck Institut f\"ur Astrophysik, Karl-Schwarzschild-Stra\ss e~1, D-85748 Garching b. M\"unchen, Germany}

\maketitle

\begin{abstract}
  Estimates of galaxy merger rates based on counts of close pairs typically
  assume that most of the observed systems will merge within a few hundred Myr
  (for projected pair separations $\leq 25h^{-1}$ kpc).  Here we investigate
  these assumptions using virtual galaxy catalogues derived from the
  Millennium Simulation, a very large N-body simulation of structure formation
  in the concordance $\Lambda$CDM cosmology. These catalogues have been shown
  to be at least roughly consistent with a wide range of properties of the
  observed galaxy population at both low and high redshift. Here we show that
  they also predict close pair abundances at low redshift which agree with
  those observed.  They thus embed a realistic and realistically evolving
  galaxy population within the standard structure formation paradigm, and so
  are well-suited to calibrate the relation between close galaxy pairs and
  mergers.  We show that observational methods, when applied to our mock
  galaxy surveys, do indeed identify pairs which are physically close and due
  to merge. The sample-averaged merging time depends only weakly on the
  stellar mass and redshift of the pair.  At $z\leq2$ this time-scale is $T
  \approx T_o r_{25} M_*^{-0.3}$, where $r_{25}$ is the maximum projected
  separation of the pair sample in units of $25h^{-1}$kpc, $M_*$ is the
  typical stellar mass of the pairs in units of $3\times
  10^{10}h^{-1}$M$_\odot$, and the coefficient $T_0$ is 1.1 Gyr for samples
  selected to have line-of-sight velocity difference smaller than 300 km/s and
  1.6 Gyr for samples where this velocity difference is effectively
  unconstrained.  These timescales increase slightly with redshift and are
  longer than assumed in most observational studies, implying that merger
  rates have typically been overestimated.\\

\end{abstract}


\section{Introduction}
\label{secc3:intro}

Ever since the pioneering work of \citet{Holmberg1937} the study of close
pairs has been considered an important tool for understanding galaxies.  Early
work was primarily directed towards comparing properties such as luminosity,
colour, and morphology with those of isolated systems, but also recognised
that the dynamics of close pairs can be used to estimate their masses
\citep[e.g.][]{Page1952}. Close pairs seemed a natural key to understanding
the initially speculative idea that galaxies might frequently merge. This was
first championed by \citet{Toomre1972} in their famous study of the dynamics
of interacting spiral galaxies, and as it was gradually accepted, mergers came
to be seen as an important factor shaping the observed galaxy population, in
particular, producing elliptical galaxies \citep[e.g.][]{Fall1979}. In its Cold Dark
Matter (CDM) incarnation, the hierarchical picture of structure growth gained
ascendancy throughout the 1980's and 1990's, and with it came ever more
detailed models which integrated merging into the build-up both of the stellar
masses and of the morphological properties of galaxies
\citep*[e.g.][]{Kauffmann1993}. Observational estimates of galaxy merger rates
thus became a critical test of the ideas underlying these theoretical models.

A number of studies have used close pair counts to estimate merger rates as a
function of redshift
\citep{Zepf1989,Burkey1994,Woods1995,Patton1997,LeFevre2000}. Such studies
assume that the observed pairs will merge on a rather short timescale,
provided they satisfy certain conditions that indicate that they are on a
tightly bound orbit. The inferred merger rate is inversely proportional to the
adopted timescale, so the results of such studies depend critically on
choosing the correct timescale with the correct dependence on pair properties
and on redshift.  Studies using this method have yielded a wide variety of
results, a diversity which can be attributed to differences in pair definition
and in the timescales adopted. No consistent picture has so far emerged.

A different technique which has become popular more recently is the
identification of mergers {\it a posteriori} through the disturbed morphology
of the merger remnants. An apparent advantage is that one doesn't have to make
any assumption about whether and when a merger will occur. Instead the merger
can be taken as a fact. On the other hand, one must adopt a timescale over
which the disturbed morphology remains visible, and this timescale is likely
to depend on redshift, on observing conditions, and on the detailed properties
of the merging systems. In practice it is highly uncertain. In addition, this
method requires high-resolution, high signal-to-noise imaging, and has
therefore become possible for the distant universe only in the last decade
with the advent of efficient space-borne imagers.

The most recent attempts to estimate merger rates with each of these methods
\citep{Lin2004,Lotz2006,Bell2006b} have indicated that evolution with redshift
is much weaker than found in earlier observational analyses and inferred from
theoretical treatments of the merging of dark-matter halos \citep[e.g.]
[]{lacey1993,Khochfar2001}. \citet{Berrier2006} gave an possible explanation
for this discrepancy based on halo-occupation-distribution (HOD) modelling of
galaxy clustering. They concluded that the galaxy merger rate does not mirror
the halo merger rate because it is strongly affected by the additional
processes which govern the merging of galaxies within a common halo. This was
demonstrated explicitly by \citet{Guo2008} using the Millennium Simulation
galaxy catalogues we analyse below. They found that whereas specific merger
rates for dark halos depend weakly on mass and strongly on redshift, the
opposite is true for galaxies, at least for the particular galaxy formation
model they analysed.

In the current paper, our focus is not on understanding these theoretical
issues, but rather on checking the assumptions which are made when estimating
galaxy merger rates from counts of close pairs. In particular, we calibrate
the relevant timescales as a function of pair properties and of redshift. We
identify close pairs in virtual galaxy catalogues following standard
observatonal criteria, and we study whether and when these pairs merge. The
simulated galaxies are embedded in a dynamically consistent way within a
realisation of the concordance $\Lambda$CDM cosmology.  Furthermore, their
properties and their small-scale clustering are a reasonably good match to
observation. Thus, we believe that the relation between close pairs and
mergers in the simulation should be similar to that in the real universe.

Our paper is organised as follows. In \secref{secc3:model} we summarise the
properties of the Millennium Simulation \citep{Springel2005b} and the
associated galaxy catalogues that we analyse here. The latter are based on the
fiducial model of \citet{Croton2006} as modified by \citet{Delucia2006b}
and extended by \citet{Kitzbichler2007}. We describe the treatment of galaxy
mergers in this model and the connection between close galaxy pairs and
mergers. We also contrast the behaviour of galaxy and halo merger rates.
\secref{secc3:methods} then explains the techniques we use to identify close
pairs and to correct for contamination by random projections. In
\secref{secc3:results} we calibrate the timescale which relates pair counts to
merger rates.  Finally the results are discussed and summarised in
\secref{secc3:discussion}.

\section{Model}
\label{secc3:model}

\subsection{The Millennium N-body simulation}
\label{secc3:millennium}
We make use of the Millennium Simulation, a very large simulation which
follows the hierarchical growth of dark matter structures from redshift
$z=127$ to the present. The simulation assumes the concordance $\Lambda$CDM
cosmology and follows the trajectories of $2160^3\sim 10^{10}$ particles in a
periodic box 500\mpch\ on a side, using a special reduced-memory version of
the {\small GADGET-2} code \citep{Springel2001b, Springel2005Gadget2}.  A full
description is given by \citet{Springel2005b}; here we summarise the main
characteristics of the simulation.

The adopted cosmological parameter values are consistent with a combined
analysis of the 2dFGRS \citep{Colless2001} and the first-year WMAP data
\citep{Spergel2003,Seljak2005}. Specifically, the simulation takes
$\Omega_{\rm m}= \Omega_{\rm dm}+\Omega_{\rm b}=0.25$, $\Omega_{\rm b}=0.045$,
$h=0.73$, $\Omega_\Lambda=0.75$, $n=1$, and $\sigma_8=0.9$, where all
parameters are defined in the standard way.  The adopted particle number and
simulation volume imply a particle mass of $8.6\times 10^8\,h^{-1}{\rm
M}_{\odot}$. This mass resolution is sufficient to represent haloes hosting
galaxies as faint as $0.1\,L_\star$ with at least $\sim 100$ particles. The
short-range gravitational force law is softened on a comoving scale of
$5\,h^{-1}{\rm kpc}$ which may be taken as the spatial resolution limit of the
calculation. The effective dynamic range is thus $10^5$ in spatial scale. Data
from the simulation were stored at 63 epochs spaced approximately
logarithmically in time at early times and approximately linearly in time at
late times (with $\Delta t \sim 300$Myr).  Post-processing software identified
all resolved dark haloes and their subhaloes in each of these outputs and then
linked them together between neighboring outputs to construct a detailed
formation tree for every halo (and its substructure) present at the final
time. The formation and evolution of the galaxy population is then simulated
in post-processing using this stored halo merger tree, as described in the
following subsection.

\subsection{The semi-analytic model}
\label{secc3:sam}

Our semi-analytic model is that of \citet{Croton2006} as updated by
\citet{Delucia2006b} and made public on the Millennium Simulation data
download site\footnote{http://www.mpa-garching.mpg.de/millennium; see
\citet{Lemson2006}\label{DBfoot}}. These models include the physical processes
and modelling techniques originally introduced by
\citet{White1991,Kauffmann1993,Kauffmann1998,Kauffmann1999,
Kauffmann2000,Springel2001} and \citet{Delucia2004}, principally gas cooling,
star formation, chemical and hydrodynamic feedback from supernovae, stellar
population synthesis modelling of photometric evolution and growth of
supermassive black holes by accretion and merging. They also include a
treatment \citep[based on that of][]{Kravtsov2004} of the suppression of
infall onto dwarf galaxies as consequence of reionisation heating. More
importantly, they include an entirely new treatment of ``radio mode'' feedback
from galaxies at the centres of groups and clusters containing a static hot
gas atmosphere. The equations specifying the various aspects of the model and
the specific parameter choices made are listed in \citet{Croton2006} and
\citet{Delucia2006b}. The only change made here is in the dust model as
described in \citet{Kitzbichler2007}.

We note that most of the assumptions made for the semi-analytic model only
affect our merger rate study in an indirect way by influencing how merging
systems are identified with observed galaxies. The dynamics of the underlying
distribution of dark matter haloes and subhaloes is not changed in any way
by the galaxy formation modelling. Only when the subhalo which hosts a galaxy
is tidally disrupted near the centre of a more massive halo does the galaxy
become eligible to merge with the central galaxy of that halo. The merger does
not occur immediately, but rather after a ``dynamical friction time''
estimated, following \citet{Binney1987}, from the relative orbit of the two
objects at the moment of subhalo disruption:
\begin{equation}
  t_{\rm fric}=1.17\frac{V_{\rm vir}r_{\rm sat}^2}{{\rm G}m_{\rm
  sat}\ln\Lambda} ~,
\label{merging_time}
\end{equation}
where $m_{\rm sat}$ and $r_{\rm sat}$ are the satellite subhalo mass
and halo-centric distance respectively, and the Coulomb logarithm is
approximated by $\ln\Lambda=\ln(1+M_{\rm vir}/m_{\rm sat})$. This
difference between the merger trees of galaxies and those of haloes
(which are assumed to merge at the instant of subhalo disruption) is
necessary since (sub)haloes can be identified only down to a certain
mass threshold.  Depending on the masses of the host and satellite
subhalos, the subhalo finder typically loses track of a subhalo when
tidal stripping has reduced its mass and dynamical friction has shrunk
its orbit to the point where it can no longer be distinguished as a
self-bound overdensity within the larger system. It is then considered
to be disrupted.  This typically occurs at radius $R \ge 1/10\, R_{\rm
vir}$, even for initially massive satellites. This is substantially
greater than the separations from which the final galaxy merger is
expected to occur.  Thus, once the subhalo disrupts, the galaxy
evolution model waits for a time $t_{\rm fric}$ before merging its
associated galaxy into the central galaxy of the main halo. During
this period the satellite galaxy has no associated subhalo and it is
assumed to remain attached to the particle which was most strongly
bound within its last identified subhalo.\footnote{Note that in the
model of De Lucia \& Blaizot (2007) which we are using, the
coefficient in equ.~\ref{merging_time} was multiplied by a factor of
two. This brings its predictions into better agreement with the recent
numerical results of \citet{Boylan2008}.}

\begin{figure}
\begin{center}
\includegraphics[width=\linewidth]{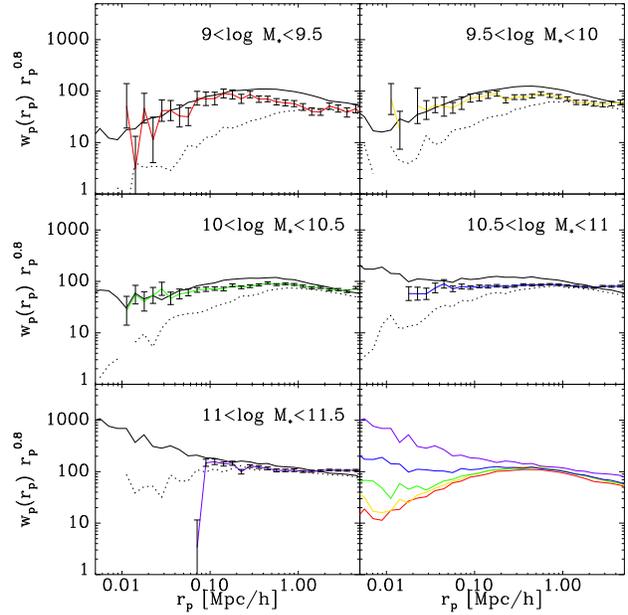}
\caption{Projected 2-point correlation function $w_p(r_p)$ for five disjoint
  stellar mass ranges both including (solid) and excluding (dashed) galaxies
  that have lost their DM subhalo. We have multiplied $w_p(r_p)$ by
  $r_p^{0.8}$ in order to reduce the dynamic range of the plots and highlight
  differences between models and observation. In the bottom right panel the
  simulation results (including ``orphan galaxies'') for all five mass ranges
  are superposed. Stellar mass increases with colour from red to purple. The
  symbols with error bars are data from the SDSS survey taken from
  \citet{Li2006}.
  \label{2pcf}}
\end{center}
\end{figure}

We can demonstrate that this treatment is required to obtain a realistic
population of close pairs by comparing the two-point correlations of our
simulated galaxies to those measured for real galaxies on scales
$r_p<100\kpch$. Such a test is presented in \fig\ref{2pcf}, which compares the
projected 2-point correlation function $w_p(r_p)$ at $z=0$ to those derived
from the SDSS survey by \citet{Li2006} for five disjoint ranges of stellar
mass. The solid black lines denote results from the simulation including all
galaxies whereas the dotted lines exclude ``orphan'' galaxies that have
already lost their surrounding (sub)halo and so shows the correlations
expected for $t_{\rm fric}=0$.  Clearly the observations cannot be fitted on
small scales by such an instantaneous merging model. The disagreement is
particularly bad for low-mass galaxies, where $w_p(r_p)$ is underpredicted by
at least a factor of 5 at scales below $r_p<100\kpch$. Observed estimates of
merger rates are typically based on counts of pairs at separations below 50
kpc, so it is clearly critical to include the ``orphan galaxies'' when
calibrating the conversion from pair counts to merger rates. Note that the
observable abundance of close pairs, after correction for random projections,
$n_{\rm pairs}$, is straightforwardly connected to $w_p(r_p)$ through the
integral
\begin{equation}
  \label{eq:wpint}
  n_{\rm pairs}(r_l)=2\pi n^2\int_0^{r_l}w_p(r_p)\,r_p\,dr_p
\end{equation}
where $n$ is the overall mean density of galaxies of the type included in the
pair sample and $r_l$ is the limiting projected separation for which pairs are
counted.

\subsection{Merger rates and pair counts}
\label{secc3:analyticpairs}

Clearly a realistic treatment of galaxy merging is crucial for our study since
we assume that the relation between simulated close pairs and simulated
mergers is a good representation of the real relation.  On the other hand, it
is important to realise that the overall merger rates in the simulation
reflect the hierarchical growth of dark halos as represented by the
halo/subhalo merger trees built from the Millennium Simulation. This determines
which galaxy pairs arrive when on the tightly bound orbits from which mergers
take place. The semi-analytic treatment of the final stages merely determines
how long each orphan--central galaxy pair ``waits'' on its tightly bound orbit
before merging. For massive pairs of the kind relevant to most observational
studies of merger rate evolution, these waiting times are often short compared
to the age of the universe at the relevant redshifts.  Thus, writing the
merging rate of orphan--central pairs of any particular type as a convolution
of the rate at which they are created through subhalo disruption with the
distribution of merging times (eq.~\ref{merging_time}),
\begin{equation}
\dot{N}_{\rm merger}(t)=\int_0^\infty \dot{N}_{\rm orphan}(t-t_{\rm
  fric}) P(t_{\rm fric}) \,dt_{\rm fric},
\label{mergrateconvol}
\end{equation}
we see that if $P(t_{\rm fric})$, the distribution of dynamical friction
timescales, is confined to values smaller than the timescales on which
$\dot{N}_{\rm orphan}$ varies, then $\dot{N}_{\rm merger} \approx \dot{N}_{\rm
orphan}$ and the semianalytic treatment has no significant effect on the
merging rate. If, on the other hand, $P(t_{\rm fric})$ has a significant tail
out to and beyond the age of the universe, the two rates can differ
significantly. Since subhalos can survive for a substantial time before they
are tidally disrupted by their host, $\dot{N}_{\rm orphan}$ differs in a
similar way from the rate at which satellite--central pairs are created
through halo merging. It is this latter rate which is often taken as a
surrogate for the galaxy merger rate.

\begin{figure}
\begin{center}
\includegraphics[width=\linewidth]{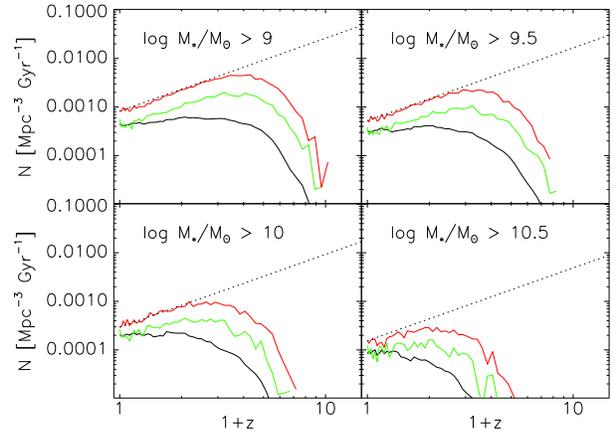}
\caption{The variation with redshift of the rate at which halo mergers create
  satellite--central galaxy pairs with stellar masses differing by less than a
  factor of four (red lines) is compared with the rate at which such pairs are
  converted to orphan--central pairs by subhalo disruption (green lines) and
  with the rate at which such pairs merge (black lines). The four panels are
  for pair samples in which the stellar masses of the individual galaxies lie
  above the four lower limits indicated. The dotted line is a powerlaw
  $\dot{N}\propto (1+z)^{1.5}$ which represents the low-redshift behaviour of
  the halo merging and satellite disruption rates, but does {\it not} fit the
  galaxy merger rate.  \label{dm_vs_gal_merg_evol_mass}}
\end{center}
\end{figure}

We illustrate these differences in Fig.~\ref{dm_vs_gal_merg_evol_mass} which
focusses on pairs of galaxies with individual stellar masses differing by less
than a factor of four and lying above the lower limits given as labels in each
panel. The red curves show the rates at which satellite--central pairs are
created by merging of their parent FOF halos. The green curves show the rate
at which corresponding orphan--central galaxy pairs are created as subhalos
disrupt, while the black curve shows the actual merger rate of these galaxy
pairs. Clearly, the delays are significant.  The orphan creation rate is a
factor of two or more below the satellite creation rate at all redshifts and
for all galaxy masses, while the galaxy merging rate is smaller again except
near $z=0$. The first difference shows that many new satellites retain their
dark matter (sub)halos for a long time. The second shows that substantial
numbers of orphan galaxies are born with relatively large $t_{\rm fric}$. Note
also that while the creation rates of satellite and orphan pairs both scale
approximately as $(1+z)^{1.5}$ at low redshift, delay effects cause the
low-$z$ galaxy merger rate to be almost independent of redshift (see below).

As we already saw in Fig.~\ref{2pcf}, at projected separations of a few tens
of kpc, counts of galaxy pairs in the Millennium Simulation are dominated by
orphan--central pairs. Thus we can approximate the abundance of observed close
pairs of any particular type as
\begin{equation}
N_{\rm close pair}(t)\approx \langle F~t_{\rm fric}\rangle~\dot{N}_{\rm
  orphan}(t) ,
\label{pairrate} 
\end{equation}
where F is a geometric factor specifying the fraction of the time a particular
orphan--central pair satisfies the observational definition of a close pair
when viewed from a random direction, the angle brackets specify an average
over all newly created pairs of the specified type, and we assume that
contributions to the average from pairs with large $t_{\rm fric}$ can be
neglected. Thus we can write,
\begin{equation}
\dot{N}_{\rm mergers}(t)\approx T^{-1} N_{\rm close pair}(t),
\label{mergerrate}
\end{equation}
where the mean timescale $T$ is defined by
\begin{equation}
T \equiv f \langle F~t_{\rm fric}\rangle, 
\label{timescale}
\end{equation}
with
\begin{equation}
f \equiv \frac{\dot{N}_{\rm orphan}}{\dot{N}_{\rm mergers}}.
\label{fdef}
\end{equation}
According to Figure~\ref{dm_vs_gal_merg_evol_mass}, the ratio $f$ increases
from 1 to about 3 as $z$ increases from 0 to 2. Equation~\ref{mergerrate} is
the standard form used to convert close pair counts to a merger rate in
observational studies.  Equation~\ref{timescale} shows how the appropriate
timescale $T$ should be estimated in the Millennium Simulation.  In practice,
we obtain it directly from the simulation data by comparing the number of
``observed'' close pairs with the merging rate. Equation~\ref{mergerrate} also
shows how the dynamical friction timescales assumed by our semi-analytic model
(equation~\ref{merging_time}) are reflected in its predictions for close pair
abundances.  The good agreement with observation in Fig.~\ref{2pcf} thus
confirms that our assumptions are realistic.  Observational studies often
assume $T\sim 500$~Myr for pair samples with projected separations below
$30\kpch$. As we will see in \secref{secc3:mergetimes}, this is an
underestimate, so the resulting merger rates are overestimates.

\subsection{Merger rates for DM haloes and galaxies}
\label{secc3:dm_vs_galaxies}

Here we digress slightly to discuss further the halo and galaxy merger rates
plotted in \fig\ref{dm_vs_gal_merg_evol_mass}.  It is immediately apparent
that all rates peak at higher redshift for smaller objects. This is because
more massive objects assemble later in hierarchical models of the kind
simulated here, and merger rates scale as the square of the abundance of the
merging population. The analytic treatment of {\it halo} mergers by
\citet{lacey1993}, based on the excursion set formalism
\citep[see][]{Press1974,Bond1991}, shows this behaviour clearly and agrees
moderately well with rates as a function of halo mass and redshift in the
Millennium Simulation; however, {\it galaxy} merger rates in the simulation
depend on stellar mass and redshift in quite a different way. For major
mergers with $M_*>10^{10}\msolh$ the galaxy merger rate depends strongly on
stellar mass but only weakly on redshift out to $z=1$, whereas the opposite is
true for dark halos \citep[see also][]{Guo2008}.

Recent observational results for galaxy mergers by \citet{Lin2004} and
\citet{Lotz2006} found a weak dependence on redshift, and these authors noted
the contradiction with theoretical predictions based on DM halo merger
rates. The contradiction was further explored by \citet{Berrier2006}, who
investigated it using HOD modelling. They inferred that the observed evolution
in merger rates requires lower halo occupation numbers at higher
redshift. This agrees with our more detailed semi-analytic treatment where it
is a consequence of the accumulation of satellite galaxies in massive host
haloes as a result their extended disruption and merging time
distributions. As is obvious from \fig\ref{2pcf}, a realistic treatment of the
accumulation requires not only the resolution of dark matter subhalos and
their associated galaxies within groups and clusters, but also a proper
treatment of orphan galaxies after their associated subhalo is disrupted.

\citet{Berrier2006} conclude that measuring galaxy merger rates is an
important tool to understand the formation and evolution of galaxies, but is a
poor probe of the cosmological aspects of structure formation; the connection
to theoretically predicted halo merger rates is subject to too many
uncertainties.  The discrepancies seen in \fig\ref{dm_vs_gal_merg_evol_mass}
support this view. On the other hand, with the advent of the {\it concordance
cosmology} most cosmological parameters appear well determined, and exploring
the details of galaxy formation is perhaps a more urgent cause. The
calibration of the galaxy merging timescale presented below accounts
realistically for differences between halo and galaxy behaviour, as judged by
the fact that the Millennium Simulation reproduces the observed clustering of
galaxies down to small scales. Nevertheless, further improvements of several
aspects of our modelling of the underlying physical processes are needed
before our calibration can be considered definitive.

\subsection{The mock lightcone}
\label{secc3:lightcone}

\begin{figure*}
\begin{center}
\includegraphics[width=\linewidth]{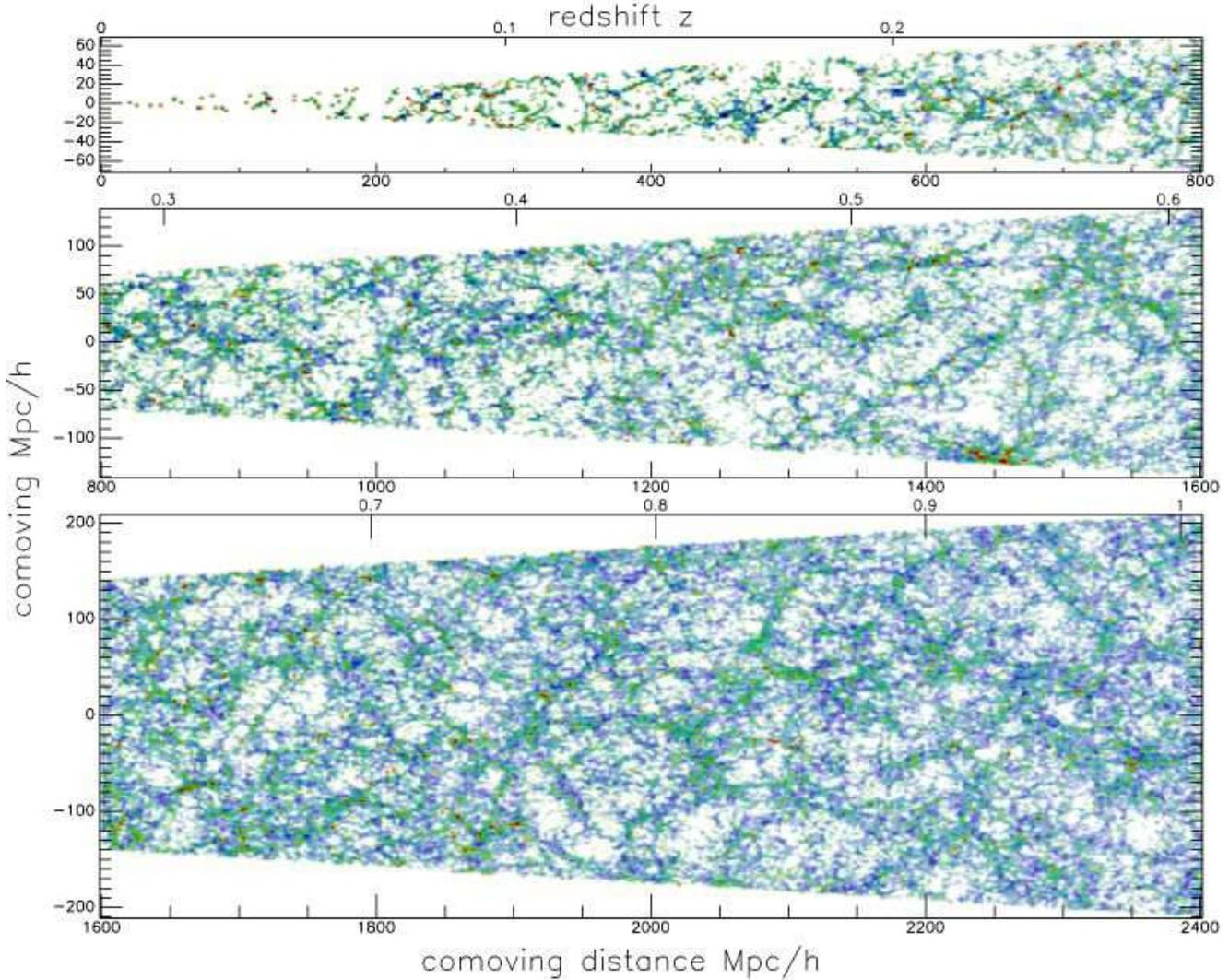}
\caption{A lightcone with a field of view of 10\,x\,1.4\,deg$^2$ which we use
  for close pair and merger rate studies. The colour map encodes projected
  galaxy density as intensity and satellite galaxy fraction as colour (from
  blue to red). Only the region out to $z=1$ is displayed, although the cone
  actually extends to $z\sim 5$.
  \label{lightcone}}
\end{center}
\end{figure*}

The fundamental question we are addressing in this paper is how well the
merger rate of galaxies can be recovered as a function of galaxy properties
from the abundance of close pairs of galaxies on the sky.  The most direct
way to assess this is to create ``mock catalogues'' from our simulation which
correspond as closely as possible to real survey catalogues, and then
to mimic observational procedures. To this end we place a virtual observer at
the origin of our simulation box and calculate which galaxies fall onto his
backward lightcone \footnote{The backward lightcone is defined as the set of
all light-like worldlines intersecting the position of the observer at
redshift zero. It is thus a three-dimensional hypersurface in four-dimensional
space-time satisfying the condition that light emitted from every point is
received by the observer now. Its space-like projection is the volume within
the observer's current particle horizon.}.  For the nearby universe these
galaxies will lie in the $z=0$ snapshot of the simulation, but as we go out
along the line-of-sight we must populate the field-of-view with galaxies from
progressively earlier snapshots. We must also interpolate redshifts, and most
importantly luminosities through various observer-frame filters, between
snapshots in order to get the appropriate values for ``observed'' properties.
A more detailed account of the methods we use to produce mock observations
from the Millennium Run semi-analytic galaxy catalogues may be found in
\citet{Kitzbichler2007}.

For the study presented in this paper we chose a field of view of
10\,x\,1.4\,deg$^2$ which we found to be a good compromise between ensuring a
sufficiently large sample for robust statistics at all redshifts of interest
and maintaining computational efficiency.  We adopt a limiting apparent
magnitude of $B_{AB}\leq 26$, close to the current effective limit for
photometric surveys of moderately large areas, and well beyond the current
limit for reliable multi-object spectroscopy. Note that because of the limited
resolution of the Millennium Simulation, our model galaxy catalogues become
incomplete at absolute magnitudes fainter than about $M_B < -16$, and as a
result our lightcone will miss intrinsically faint galaxies at all but the
highest redshifts. This will not affect our later analysis which is restricted
to bright and massive systems.

Our final mock catalogue contains 3236337 galaxies.
In \fig\ref{lightcone} we depict their spatial distribution out to
$z=1$ in order to illustrate the structure in this mock lightcone. The
filaments and voids emerge vividly in this plot, where we encode projected
galaxy density as intensity and satellite galaxy fraction as colour. Clearly 
many galaxies in the most clustered regions are satellites, whereas in
the filaments and the sparsely populated regions, most galaxies are the
central systems of their halos.

\section{Pair selection methods}
\label{secc3:methods}

\subsection{Finding pairs}
\label{secc3:pairfind}

A limitation of our mocks in comparison to real catalogues is that they
include no record of recent close interactions which might be related to the
morphological indicators accessible with high-quality deep imaging. Many
authors, beginning with \citet{Toomre1972} and \citet{Larson1978}
have shown that close encounters between massive galaxies can produce both
enhanced star formation and disturbed morphologies \citep[e.g][and references
therein]{Patton2005,Lin2006,Li2008}.  Detection of such effects is a clear
indicator that apparent proximity on the sky does indeed correspond to
physical interaction, and so greatly increases the level of confidence that a
given close pair is likely to merge. On the other hand, the detectability of
these effects depends strongly on the quality of the imaging, on the structure
of the merging galaxies, and on the time, viewing angle and redshift at which
they are observed.  As a result it is very difficult to estimate
what fraction of close pre-merger pairs will be detected by any given set of
morphological criteria. This makes it impossible to estimate merger rates
reliably from such samples.

Until recently, observational studies of merging typically involved from a few
dozen to a few hundred pairs. Every pair could be examined visually to assess
whether it is interacting. Current and future surveys will produce much larger
samples for analysis, necessitating automatic techniques to search for
morphological signatures of interaction. The reliability of such
classification techniques depends crucially on good signal-to-noise and
adequate resolution. When these conditions are met, measures of concentration,
asymmetry and clumpiness can be combined with other indices such as the {\it
Gini} and $M_{20}$ coefficients of \citet{Lotz2004} to produce very large
samples of galaxies with a morphological classification \citep[see
e.g.][]{Abraham2003,Prescott2004,Zamojski2006}, of which $1-3\%$ typically
show signatures of an ongoing interaction. For the reasons noted above,
however, such samples are not suitable for estimating merger rates.  For the
rest of this paper we will therefore concentrate on pair samples selected
purely by the proximity of the two galaxies.

\subsubsection{Pair samples from imaging alone}
The most straightforward way to find pairs of galaxies is simply to identify
objects which are close together on the sky in a purely photometric
survey. This technique was used for some of the earliest pair fraction studies
\citep[e.g.][]{Zepf1989} because it could be applied to any survey with a
large enough galaxy catalogue ($>1000$ at that time). One must keep in mind
that the close pair fraction is of order a few percent, so to get acceptable
statistics for the pair sample, the original catalogue must be much
larger. The disadvantage of this purely photometric method is, of course, that
one will inadvertently include many false pairs, i.e. chance projections that
are not physically close.  This ``background noise'' becomes more problematic
for higher mean galaxy densities on the sky, corresponding to deeper magnitude
limits; early studies worked moderately well {\it because} of their shallow
limits.

The fraction $F$ of true companions in a sample of apparent pairs can be
estimated from the angular correlation function $w(\theta)$ as
$F=w/(1+w)$. Only for $w(\theta)>1$ are the majority of apparent
companions at angular distance $\theta$ true physical companions. According to
Limber's equation \citep{Limber1953} the angular two-point correlation
function depends on limiting flux density $f=\nicefrac{L}{4\pi r^2}$ as
$w(\theta)\propto f^{\gamma/2}$ (assuming a power law $\xi=(r_0/r)^\gamma$ for
the spatial function with $r_0$ independent of distance).  For surveys as deep
as we simulate here, $w(\theta)\sim 1$ corresponds to $\theta\sim 0.1$arcsec
so that observationally realistic samples of close pairs (typically limited to
separations of a few arcseconds) are entirely dominated by chance
projections. Although for large samples the fraction of ``true'' close pairs
can be determined statistically with high reliability, it is impossible to
know {\it which} close pairs are interacting without additional information,
for example from morphologies. Furthermore, without spectroscopy the
separation distribution (in 3-D) of the true pairs and its dependence on
redshift cannot be derived from the observed angular separation distribution
without making additional assumptions about the redshift distribution of the
population and the evolution of its clustering.

\subsubsection{Primary redshift catalogue with photometric companions}
Many recent pair studies \citep[e.g.][]{Yee1995} have been based on
correlating a redshift survey with a deeper photometric catalogue.  This
allows the identification of all close apparent companions for a complete set
of galaxies of known distance and brightness. For sufficiently large samples
the projected correlation function $w_p(r_p,z)$ can be estimated, giving the
abundance of true physical pairs as a function of projected separation $r_p$.
Assuming isotropy of orientation for the underlying population, this can be
inverted to give the distribution of companions as a function of 3-D
separation, and thus the abundance of companions within some maximal
separation (e.g. 30 kpc). Note that without morphological information one
still has no indication of {\it which} apparent pairs are actually physically
close.  This problem is significant in deep surveys where the majority of
apparent projections are chance superpositions of unrelated objects. The major
advantage of starting with a redshift survey is that the dependences of the
close pair distribution on physical separation and on redshift can be
determined separately.

\subsubsection{Photometric redshift pair identification}
If photometric redshifts are available for all galaxies in a catalogue, this
allows the definition of still purer samples of physical pairs. Here also one
can define a physical (rather than angular) search radius around each galaxy,
and additionally one can limit acceptable pairs to those whose redshifts are
equal to within the accuracy of the photometric determinations. Some
correction for random pairs is still required, however, since this accuracy is
sufficiently poor that projected pairs with moderately large true redshift
differences can still enter the sample. The number of ``true'' pairs at any
given apparent separation $r_p$ and redshift $z$ within some photometric
redshift tolerance $\Delta z$ can be found by taking the number of such pairs
counted in the real catalogue and subtracting the mean number found in a large
number of artificial catalogues in which the photometric redshifts of the
galaxies are retained but their angular positions within the survey area are
randomised.

\subsubsection{Complete spectroscopic redshift samples}

Clearly the ideal sample for a pair study is one that includes accurate
spectroscopic redshifts for all galaxies. This allows a search for ``true''
physical companions in the space of projected physical separation and velocity
difference. The result is an unbiased sample with minimal contamination by
optical pairs. In principle, a correction for random pairs can be applied just
as in the previous section, but in practice this correction is so small that
it can be neglected.  Additionally, one can estimate the fraction of the
physical pair population which corresponds to truly close pairs, i.e. to pairs
for which the 3-D separation is also small. 

\subsection{Identifying candidate pairs for mergers}
\label{secc3:mergedef}

From our mock survey lightcone we construct several close pair samples as
follows. For each galaxy we examine the 20 closest companions on the sky and
apply various criteria to define pair subsets that we consider as merger
candidates. These criteria include: (i) projected physical separation $r_p$,
(ii) radial velocity difference $\Delta v$, (iii) redshift difference $\Delta
z$. We apply these cuts in different combinations to build different
samples. In addition, we distinguish pairs by the stellar mass ratio of the
two pair members.

For the rest of this paper we will concentrate on potential major mergers
which we define to be pairs with stellar mass ratios of \mbox{4:1} or
less. This restriction is applied for several reasons. First, observational
studies usually concentrate on galaxy pairs with small magnitude differences,
either because both galaxies are typically close to the apparent magnitude
limit of the parent survey, or because a limit on apparent magnitude
difference is applied explicitly. This is to prevent confusion between actual
companions and morphological features in the outer regions of a bright galaxy.
Restricting galaxy pairs to a narrow range of mass ratios also makes sense
from a theoretical point of view, since it is the growth of galaxies through
major mergers that dominates the morphological transformation of galaxies.

Using the criteria listed above we define a number of samples. For the
projected physical (i.e. {\it not} comoving) distance $r_p$ we choose maximal
values of 30, 50, or 100$\kpch$.  To mimic ``spectroscopic'' samples, we
assume infinitely accurate redshifts and select pairs with radial velocity
differences $\Delta v<300\kms$. (Note that this excludes a number of true
physical pairs with larger velocity separation, but most such pairs are within
massive clusters and so rarely merge.)  For ``photo-z'' samples we require a
redshift difference of $\Delta z<0.05$. In the following sections we will use
pair samples defined in this way to study the relation between close pairs of
galaxies and mergers.

\section{Results}
\label{secc3:results}

\begin{figure}
\begin{center}
\includegraphics[width=\linewidth]{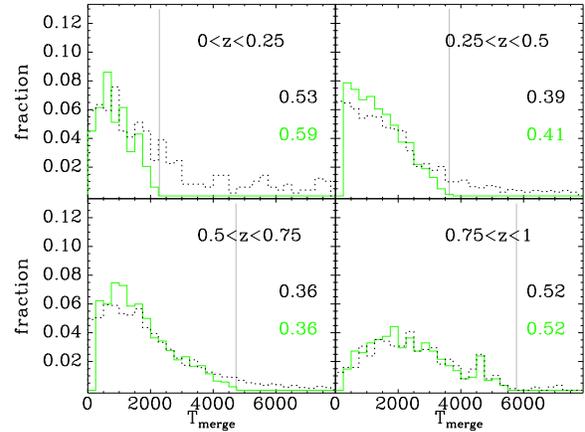}
\caption{Distribution of merging times for galaxies more massive than
  $M_*>10^{10}\msolh$ at four different redshifts. Pairs are selected from the
  lightcone with $r_p<50\kpch$ and $\Delta v<300\kms$. For the green
  histograms, merging times were determined by following the galaxies forward
  in time until they merge (or reach $z=0$). The timespan between the highest
  redshift contributing to each panel and $z=0$ is indicated by the grey
  vertical line. For the black histograms, merging times were determined using
  the internal counters set when one of the galaxies first loses its dark
  halo. (This can occur before or after the pair is actually identified in the
  lightcone.) All samples are subject to an apparent magnitude cut at $B<26$
  and only major mergers are considered. The coloured numbers in each panel
  give the fraction of all pairs which are not predicted to merge by $z=0$.
\label{mergetimes}}
\end{center}
\end{figure}

\begin{figure}
\begin{center}
\includegraphics[width=\linewidth]{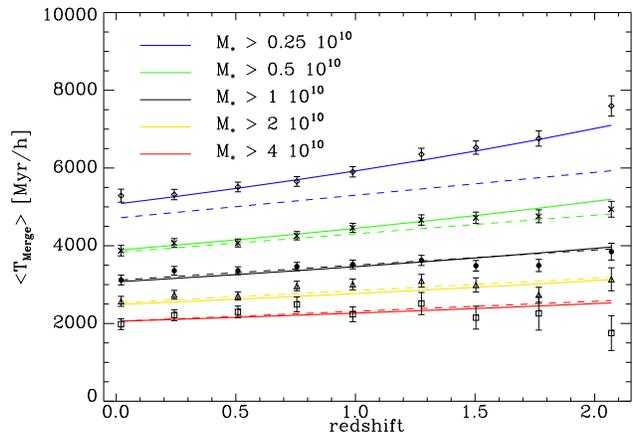}
\caption{Redshift evolution of the timescale $T=N_{\rm pairs}/\dot N_{\rm
  Merge}$ for conversion from pair fraction to merger rate.  Two-dimensional
  linear regression fits (\eq\ref{eq:timescalefit}) are plotted as solid
  curves for a range of mass cuts denoted by different colours, as indicated
  by the labels.  The corresponding data are indicated by points with error
  bars.  All pair samples in this plot were selected requiring projected
  separations $r_p<50\kpch$, radial velocity difference s$\Delta v<300\kms$,
  and galaxy stellar masses differing by a factor less than 4. The dashed
  lines are for the simplified fitting function of
  \eq\ref{eq:timescalefitsimplspec}.
\label{calib_z_evol}}
\end{center}
\end{figure}

\subsection{Distribution of merging times}

\label{secc3:mergetimes}

In \fig\ref{mergetimes} we show distributions of merging times for close pairs
of galaxies in our lightcone with $r_p<50\kpch$, $\Delta v<300\kms$,
individual apparent magnitudes $B<26$, and individual stellar masses which
exceed $10^{10}\msolh$ and differ by less than a factor of 4. The four panels
show distributions for four disjoint redshift ranges as indicated.  Merging
times were determined either by following the later evolution of each pair
until merging (or until $z=0$; the green histograms) or by using the
time-until-merger counter assigned to each orphan galaxy at the time it is
orphaned (the dashed black histogram). The distributions are plotted as the
fraction of {\it all} pairs in each histogram bin, and so do not normalise to
unity in the green case. The fraction of pairs which do {\it not} merge by
$z=0$ is indicated in each panel by labels of the appropriate colour.

The most important results to note from this figure are that the merger time
distributions vary little with redshift, that they extend to large values, and
that they include the majority of pairs. Most close pairs eventually merge,
even for $r_p< 50\kpch$. These results are best seen from the black
histograms. These indicate a median merger time above 2~Gyr, much longer than
the merging times typically adopted when estimating merger rates from observed
pair counts. At lower redshifts, the directly estimated merger-time
distributions do not extend to large times. This simply reflects the fact that
there is insufficient time for many of the mergers to take place, as may be
seen from the vertical grey lines which give the look-back time to the largest
redshift used when constructing the distributions in each panel. The black
histograms show how much longer one would have to wait for the other objects
to merge. At merger-times below this limit there is good agreement between the
directly and indirectly estimated distributions (the black and the green
histograms).

The distributions of merger-times in the highest redshift panel appear to have
fewer pairs with short merger times than those at lower redshift. This is
because the imposed apparent magnitude limit at $B>26$ excludes significant
numbers of galaxies from the sample at these redshifts. The galaxies that are
lost are primarily red systems close to our mass cut at $10^{10}\msolh$. These
are almost all satellite systems which have had substantial time to age and
dim since their accretion; they are thus typically ``about'' to merge.  This
effect is also responsible for the fact that the fraction of observed pairs
which do {\it not} merge by $z=0$ increases in the highest redshift panel,
reversing the trend in the other panels. It seems that selection effects may,
in some circumstances, bias observational samples against pre-merger pairs,
although interaction-induced star formation (which is not included in our
galaxy modelling) could well reduce or even reverse this bias.

\subsection{Mean merging times}
\label{secc3:avetime}

We have established that, for the separation and velocity difference cuts
typically adopted, most close pairs of similar mass galaxies will, in fact,
merge. We can therefore address the main issue of this paper, namely: ``What
timescale should be used to convert counts of such close pairs into a merger
rate?'' As noted in \eq\ref{mergerrate}, this timescale is simply the
ratio at each redshift of the abundance of pairs of a particular type to the
merger rate of such pairs per unit volume,
\begin{equation}
  \label{eq:timescale}
  \langle T_{\rm merge}\rangle=N_{\rm pairs}/\dot N_{\rm merge}.  
\end{equation}
Calculating this ratio as a function of redshift and mass cut for pairs with
$r_p<50\kpch$ and $\Delta v<300\kms$ yields the results presented in
\fig\ref{calib_z_evol}. Since the square root of the inverse of this
dependency seems to be linear within the scatter for mass cuts below
$10^{10}\msolh$, we decided to apply a two-dimensional linear regression to
$\langle T_{\rm merge}\rangle^{-\nicefrac{1}{2}}\equiv
T^{-\nicefrac{1}{2}}(z,M_*)$ as a function of $z$ and $\log M_*$, implying the
relation
\begin{equation}
  \label{eq:timescalefit}
  \langle T_{\rm merge}\rangle^{-\nicefrac{1}{2}}=T_0^{-\nicefrac{1}{2}}+f_1\,z+f_2\,(\log M_*-10)\,.
\end{equation}
The value of $T_0$ as well as the coefficients $f_x$ and their
uncertainties estimated from fits to all our numerical data are tabulated for
samples with different pair identification criteria in \tabl\ref{tconfacs}.

In the low redshift regime ($z\leq 1$)
and for stellar masses above $5\times 10^{9}\msolh$ an even simpler fitting
formula works well:
\begin{equation}
  \label{eq:timescalefitsimplspec}
  \langle T_{\rm merge}\rangle=2.2{\rm Gyr}\frac{r_p}{50{\rm kpc}}\ \left(\frac{M_*}{4\cdot10^{10}\msolh}\right)^{-0.3}(1+\frac{z}{8})
\end{equation}
for samples restricted to $\Delta v<300\kms$  and
\begin{equation}
  \label{eq:timescalefitsimplphot}
  \langle T_{\rm merge}\rangle=3.2{\rm Gyr}\frac{r_p}{50{\rm kpc}}\ \left(\frac{M_*}{4\cdot10^{10}\msolh}\right)^{-0.3}(1+\frac{z}{20})
\end{equation}
for samples limited to $\Delta v<3000\kms$. These simplified fits give the
results indicated by the dashed lines in \fig\ref{calib_z_evol}.  The
difference in the normalisation coefficient between the two cases reflects the
fact that expanding the velocity cut admits about 50\% more pairs.  Most of
these additional pairs are physically associated but lie within larger groups
or clusters. The timescales for $\Delta v<3000\kms$ should be used when
analysing data from photometric redshift samples, since the ``background''
correction will not eliminate physically associated galaxies at large velocity
separation.

Aside from the dependence on mass cut and redshift that is illustrated in this
figure, there is also a strong dependence on the maximum projected radius
$r_p$. This is a natural consequence of \eq\ref{eq:timescale},
since the denominator $\dot N_{\rm merge}$ is independent of $r_p$ whereas the
numerator $N_{\rm pairs}$ is not. The latter is proportional to the integral
of the projected \mbox{2-point} correlation function $w_p(r)$ out to $r_p$
(see \eq\ref{eq:wpint}). If we choose the usual parametrisation $w_p\sim
(r/r_0)^{-\alpha}$ we get $N_{\rm pairs}\sim r_p^{2-\alpha}$, where
$\alpha=0.8$ is commonly adopted in the literature.  Thus we would expect the
values in the table to scale as $r_p^{1.2}$ which is qualitatively consistent
with the actual values but slightly too strong. If we instead calculate
$\alpha$ from the measured values, we get $\alpha=1.06$ and $\alpha=0.93$ for
the intervals \mbox{30-50$\kpch$} and \mbox{50-100$\kpch$} respectively.  This
is consistent with \fig\ref{2pcf} where we see that the projected
\mbox{2-point} correlation function on scales below $100\kpch$ and for masses
above $3\cdot10^{10}\msolh$ is considerably steeper than the fiducial
$\alpha=0.8$.

The mean merging times found here are clearly consistent with the distribution
of individual merging times shown in \fig\ref{mergetimes}.  They are also much
larger than the values $\sim 0.5$~Gyr typically adopted in observational
studies of this problem.  As a result most earlier studies have
substantially overestimated merger rates.

\begin{table}
  \caption{Coefficients for different pair identification criteria obtained
    from fits of $\langle T_{\rm merge}\rangle=T(z,M_*)$ to our numerical data
    on $N_{\rm pairs}/\dot N_{\rm merge}$ according to
    \eq\ref{eq:timescalefit}.
    \label{tconfacs}}
\begin{center}
\input{regrtable_rp_vp.tex}
\end{center}
\end{table}

\section{Conclusions}
\label{secc3:discussion}

We have investigated major merger rates in our semi-analytic model
based on the Millennium N-body simulation and compared them to the
abundance of close galaxy pairs. In this way we have calibrated the
relation used to estimate merger rates from deep galaxy surveys. In
addition, we have shown that for the parameters typically adopted in
observational studies, most close pairs do indeed merge, albeit on a
substantially longer timescale than is usually assumed. As a result,
the characteristic timescales we derive are indeed the typical times
until pair members merge. The ideal parent catalogue for such studies
would contain spectroscopic redshifts for all galaxies, but in
practice reliable results can be obtained from any deep photometric
catalogue, provided good photometric redshifts are available and care
is taken to correct for chance line-of-sight projections. The main
advantage of using photo-$z$'s is, of course, that they allow results
to be obtained for much larger and deeper samples than could otherwise
be used. Their main disadvantage is that one does not know which close
pairs are ``physical'' and which are random projections.

\begin{enumerate}
\item[] The main results of our study are as follows:
\item The characteristic timescale which
  converts background-corrected pair counts into merger rates
  (\fig\ref{calib_z_evol}) depends on the pair identification criteria,
  on the stellar mass cut and weakly on the redshift. For stellar masses above
  $5\times 10^{9}\msolh$ it can be approximated by the simple relations
  $$\langle T_{\rm merge}\rangle=2.2{\rm Gyr}\frac{r_p}{50{\rm kpc}}\ \left(\frac{M_*}{4\cdot10^{10}\msolh}\right)^{-0.3}(1+\frac{z}{8})$$
  for radial velocity differences $\Delta v<300\kms$ and by
  $$\langle T_{\rm merge}\rangle=3.2{\rm Gyr}\frac{r_p}{50{\rm kpc}}\ \left(\frac{M_*}{4\cdot10^{10}\msolh}\right)^{-0.3}(1+\frac{z}{20})$$ 
  for $\Delta v<3000\kms$. This latter relation should be used for pair counts
  derived from photometric redshift surveys.  A more accurate fitting formula
  is given in \eq\ref{eq:timescalefit}; the corresponding coefficients $T_0$,
  $f_1$ and $f_2$ are listed in \tabl\ref{tconfacs} for a range of pair
  selection criteria.
\item The characteristic timescales we find are larger (typically by a
  factor of at least 2) than is assumed in most published determinations of merger
  rates. These are therefore likely to be substantial overestimates of the
  true rates.
\item For masses $M_*>3\times 10^{9}\msolh$, the intrinsic galaxy merger rate
  evolution is quite flat at low redshift, $\dot N\sim(1+z)^\alpha$, with
  $\alpha < 0.5$ and decreasing towards higher mass. For large masses the
  exponent becomes negative. Overall, the distributions are quite flat out
  to redshift \mbox{$z\sim 2$} (see e.g. \fig\ref{dm_vs_gal_merg_evol_mass}).
  Observational results lie in the range $N_{\rm pair}\sim(1+z)^{2\pm2}$ where
  the large uncertainties are presumably due to small sample sizes and
  uncontrolled selection effects.  In particular, effects due to the apparent
  magnitude limits of real surveys interact with the stellar populations of
  galaxies in ways which make it very difficult to define physically
  equivalent samples at different redshifts.  We have presented most of our
  results for volume-limited samples in order to avoid confusion due to these
  complexities.
\item The broad distribution of merging times, peaking well beyond
  1~Gyr, results in merger rates for galaxies which evolve differently
  from those of dark matter halos, even of halos similar in mass to those that
  host galaxies. At low redshifts merger rates for DM halos scale as $\dot
  N\sim(1+z)^{\nicefrac{3}{2}}$ for all masses, a much more rapid evolution
  than we find for galaxies (\fig\ref{dm_vs_gal_merg_evol_mass}). This
  discrepancy has already been described by other authors, and we agree with
  their conclusion that merger rate studies are less suitable for probing the
  overall growth of cosmic structure than originally thought.  They can instead
  contribute substantially to our understanding of the formation and evolution
  of galaxies.
\end{enumerate}

{\bf Acknowledgements:} MGK acknowledges a PhD fellowship from the
International Max Planck Research School in Astrophysics, and support from a
Marie Curie Host Fellowship for Early Stage Research Training.

\bibliographystyle{aa}
\bibliography{references}

\end{document}

%% file: regrtable_rp_vp.tex
\begin{tabular}{crrr}
\hline\hline
{\scshape Velocity} & \multicolumn{3}{c}{\scshape Projected distance}\\
\hline
 & \multicolumn{3}{c}{$r_p$} \\
{\raisebox{1.5ex}[-1.5ex]{$v_p<300\kms$}} & {$\le30\kpchh$} & {$\le50\kpchh$} & {$\le100\kpchh$}\\
\hline
$T_0 [\myrh]$\dotfill  & 2038 & 3310 & 6909 \\
$10^{5}f_1 [\myrh^{\nicefrac{-1}{2}}]$\dotfill  & $ -165.\pm   4.4$ & $ -105.\pm   3.3$ & $ -30.4\pm   2.2$ \\
$10^{5}f_2 [\myrh^{\nicefrac{-1}{2}}]$\dotfill  & $  690.\pm   10.$ & $  668.\pm   7.7$ & $  571.\pm   5.2$ \\
\hline
 & \multicolumn{3}{c}{$r_p$} \\
{\raisebox{1.5ex}[-1.5ex]{$v_p<3000\kms$}} & {$\le30\kpchh$} & {$\le50\kpchh$} & {$\le100\kpchh$}\\
\hline
$T_0 [\myrh]$\dotfill  & 2806 & 4971 & 11412 \\
$10^{5}f_1 [\myrh^{\nicefrac{-1}{2}}]$\dotfill  & $ -94.7\pm   3.7$ & $ -38.6\pm   2.7$ & $  18.0\pm   1.7$ \\
$10^{5}f_2 [\myrh^{\nicefrac{-1}{2}}]$\dotfill  & $  671.\pm   8.7$ & $  615.\pm   6.3$ & $  491.\pm   4.2$ \\
\hline
\end{tabular}